\documentclass[letter]{aa}

\usepackage{graphicx}
\usepackage{txfonts}

\begin{document}

\title{Detection of sea-serpent field lines in sunspot penumbrae}
\author{A.\ Sainz Dalda\inst{1,2} \and L.R.\ Bellot Rubio\inst{3}}
\institute{THEMIS S.L., C/V\'{\i}a L\'actea s/n, 38200 La Laguna, 
Tenerife, Spain \and 
Instituto de Astrof\'{\i}sica de Canarias , C/V\'{\i}a L\'actea s/n, 
38200 La Laguna, Tenerife, Spain \and
Instituto de Astrof\'{\i}sica de Andaluc\'{\i}a, CSIC, Apdo.\ 3004, 
18080 Granada, Spain}
\date{Received / Accepted }

\abstract{} {We investigate the spatial distribution of magnetic
polarities in the penumbra of a spot observed very close to disk
center.}  
{High angular and temporal resolution magnetograms taken
with the Narrowband Filter Imager aboard Hinode are used in this
study.  They provide continuous and stable measurements in the
photospheric \ion{Fe}{i} 630.25 line for long periods of time.}
 {Our observations show small-scale, elongated, bipolar magnetic
structures that appear in the mid penumbra and move radially
outward. They occur in between the more vertical fields of the
penumbra, and can be associated with the horizontal fields that harbor
the Evershed flow. Many of them cross the outer penumbral boundary,
becoming moving magnetic features in the sunspot moat. We determine
the properties of these structures, including their sizes, proper
motions, footpoint separation, and lifetimes.}  
{The bipolar patches can be interpreted as being produced
by sea-serpent field lines that originate in the mid penumbra and
eventually leave the spot in the form moving magnetic features.  The
existence of such field lines has been inferred from Stokes inversions
of spectropolarimetric measurements at lower angular resolution, but
this is the first time they are imaged directly. Our observations add
another piece of evidence in favor of the uncombed structure of
penumbral magnetic fields. }

\keywords{Sunspots -- Sun: magnetic fields -- Sun: photosphere --
Magnetohydrodynamics (MHD)}

\titlerunning{Detection of sea-serpent field lines in sunspot penumbrae}
\authorrunning{Sainz Dalda and Bellot Rubio}
\maketitle

\section{Introduction} 

Significant progress in the understanding of the penumbra has
been made through high resolution magnetograms and Dopplergrams 
(Title et al.\ 1993; Schlichenmaier \& Schmidt 2000; Langhans et al.\
2005; Rimmele \& Marino 2006), spectropolarimetric measurements 
(Degenhardt \& Wiehr 1991; S\'anchez Almeida \& Lites 1992; Lites et
al.\ 1993; Stanchfield et al.\ 1997; R{\"u}edi et al.\ 1999;
Westendorp Plaza et al.\ 2001; Schlichenmaier \& Collados 2002; Mathew
et al.\ 2003; Bellot Rubio et al.\ 2004; Borrero et al.\ 2005, 2006;
S\'anchez Cuberes et al.\ 2005; Beck 2006; Jur\v{c}\'ak et al.\ 2007),
and forward modeling (Mart\'{\i}nez Pillet 2000; M{\"u}ller et
al.\ 2002). Today it is agreed that the penumbra consists of magnetic
fields having different inclinations and strengths, as proposed by
Solanki \& Montavon (1993) in their uncombed penumbral model. One
particularly relevant result derived from spectropolarimetry is that
some field lines return to the solar interior well within the
penumbra.  The modest angular resolution of ground-based polarimeters
has precluded an unambiguous direct detection of such
opposite-polarity field lines or the study of their temporal
evolution.

Also, in recent years there has been an increased interest in determining 
the relation between moving magnetic features (MMFs; Harvey \& Harvey 1973) and
penumbral magnetic fields. There is growing observational evidence that
bipolar MMFs are the continuation of the more horizontal magnetic fields of
the penumbra in the sunspot moat (Sainz Dalda \& Mart\'{\i}nez Pillet 2005;
Ravindra 2006; Cabrera Solana et al.\ 2006; Kubo et al.\ 2007a), but a clear
picture has not yet emerged due to insufficient angular resolution. 

Establishing the small-scale organization of the magnetic field in the
penumbra is crucial to shed light on these issues, and also to
distinguish between competing models of the penumbra. Here we
investigate the distribution of magnetic polarities in and around
sunspots using magnetograph observations taken with
Hinode (Kosugi et al.\ 2007). The unprecedented resolution and
stability of these measurements allow us to follow the evolution of
the penumbral fine structure in polarized light for hours.

\section{Observations} 

Our study is based on longitudinal magnetograms acquired with the
Narrowband Filter Imager (NFI; Tarbell et al.\ 2008) aboard Hinode.
On November 14, 2006, the NFI observed the isolated spot AR 10923 from
07:10 to 09:40 UT and from 11:00 to 17:10 UT.  The instrument was
tuned to measure the Stokes $I$ and $V$ signals of \ion{Fe}{i}
630.25~nm at $-120$~m\AA\/ from line center. The magnetograms cover 
a field of view (FOV) of $328\arcsec\/ \times 164\arcsec$, and have an
irregular cadence of 1-5 minutes. The effective pixel size of the
measurements is 0\farcs16, which gives a spatial resolution of about
0\farcs32.  The spectral resolution of the NFI is 90 m\AA\/. The 
spot crossed the central meridian around 10 UT, reaching a minimum
heliocentric angle of 6$^\circ$.

The data have not been corrected for instrumental effects because of
difficulties in constructing accurate flatfields. We therefore
restrict our analysis to the polarity of the magnetic field.  The
polarity should be insensitive to spatial variations in the instrument
transmission because it is obtained as the difference of two intensity
measurements.

\begin{figure*}[p]
\centering
\includegraphics[width=17.3cm,bb=-30 0 820 702]{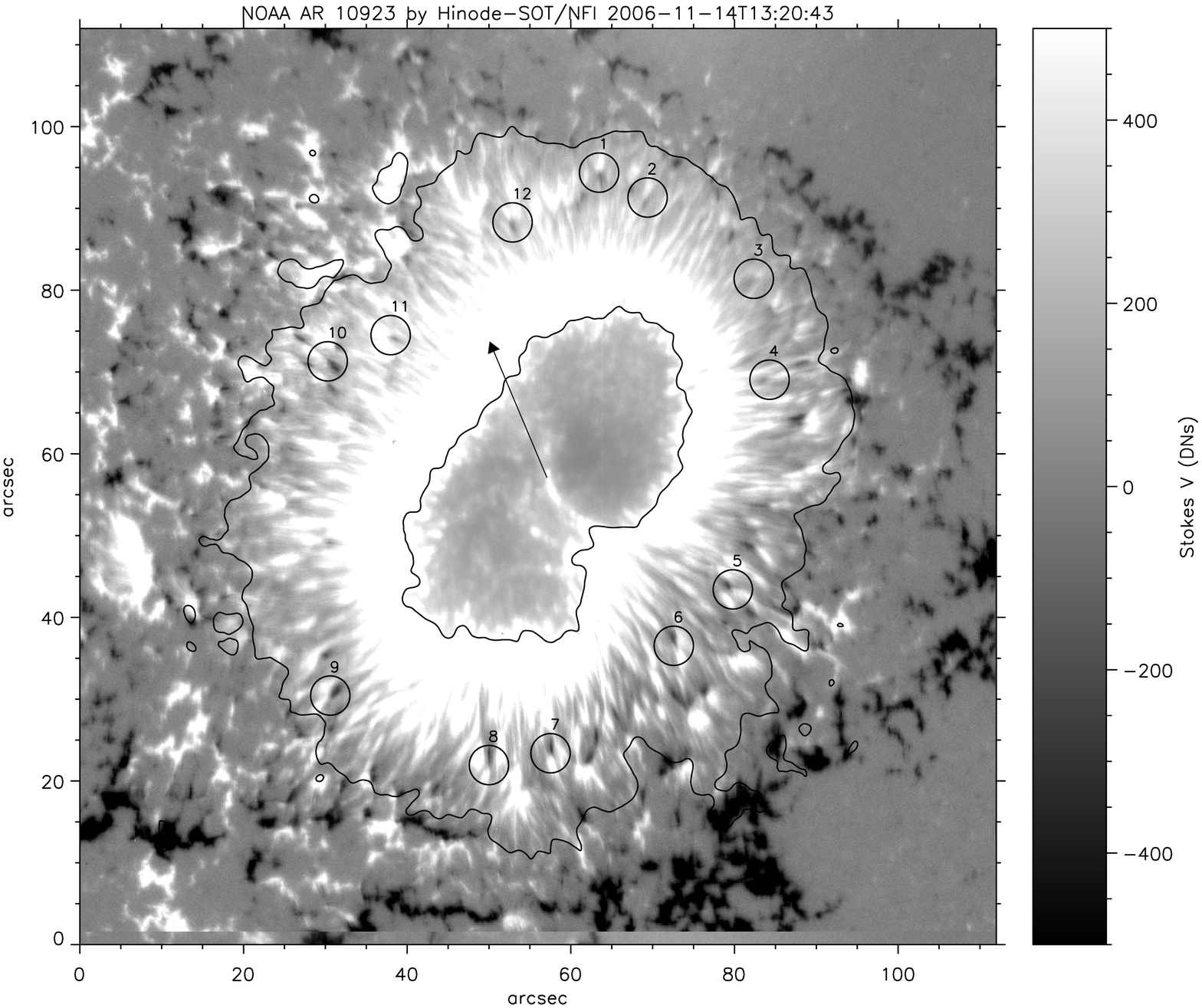}
\includegraphics[height=4.2cm]{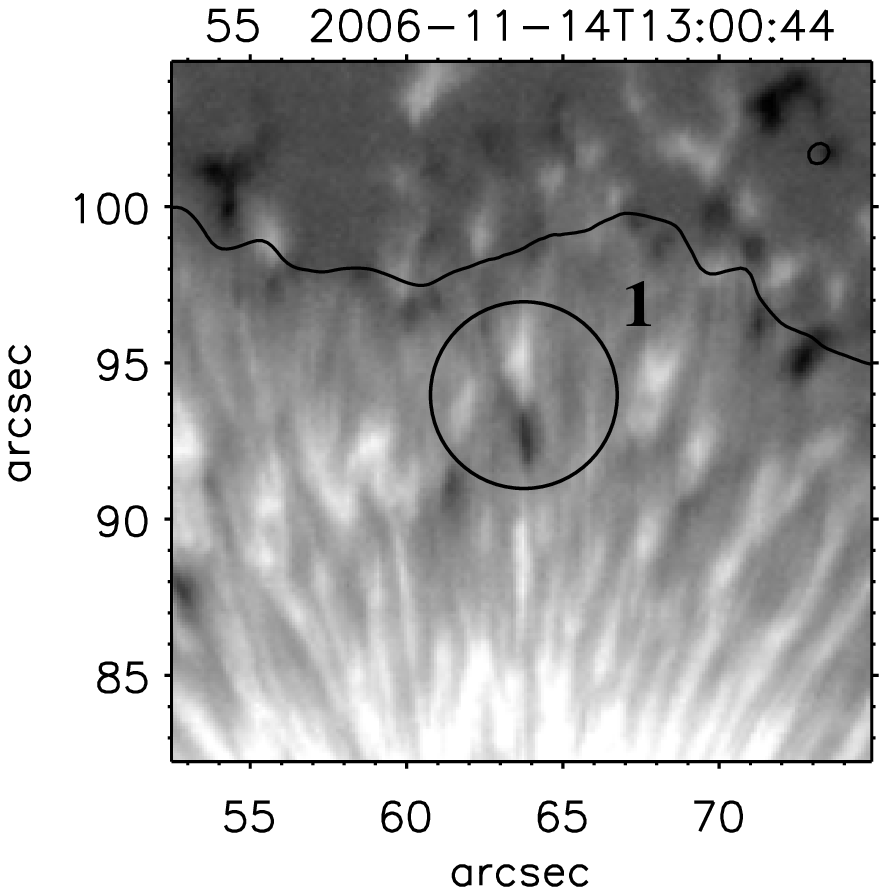}
\includegraphics[height=4.2cm]{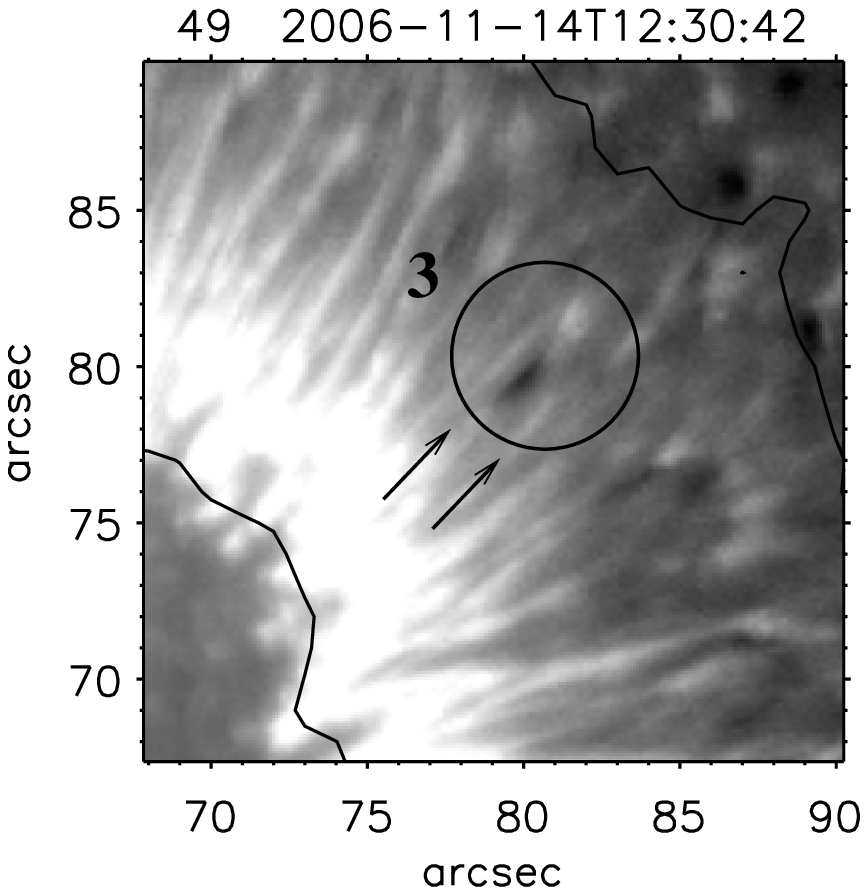}
\includegraphics[height=4.2cm]{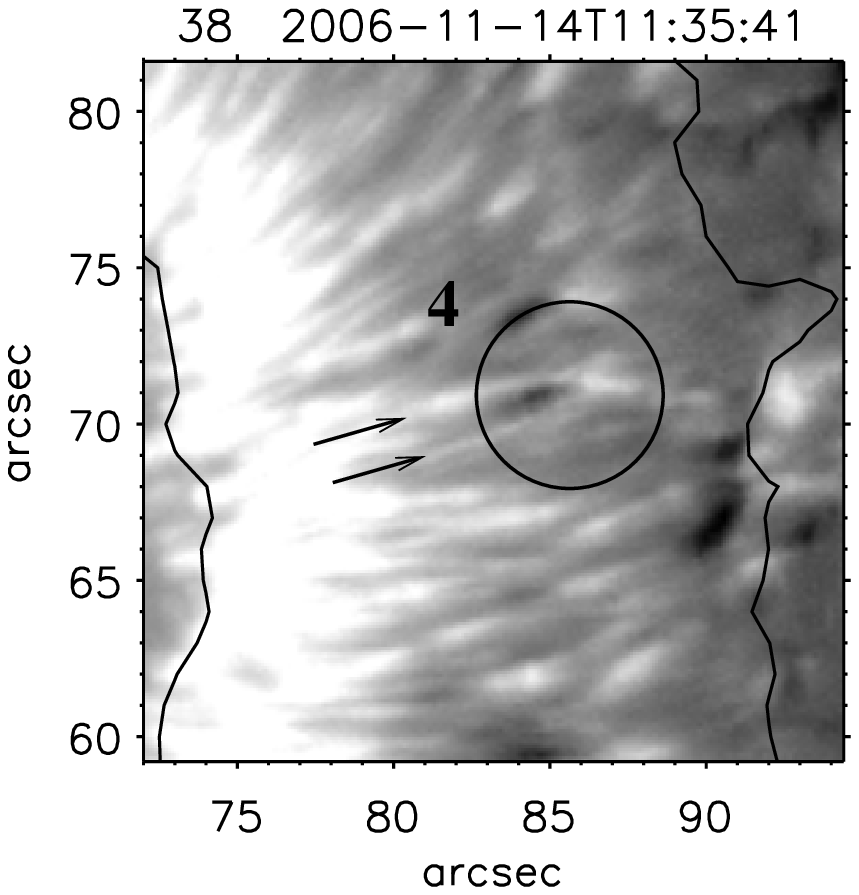}
\includegraphics[height=4.2cm]{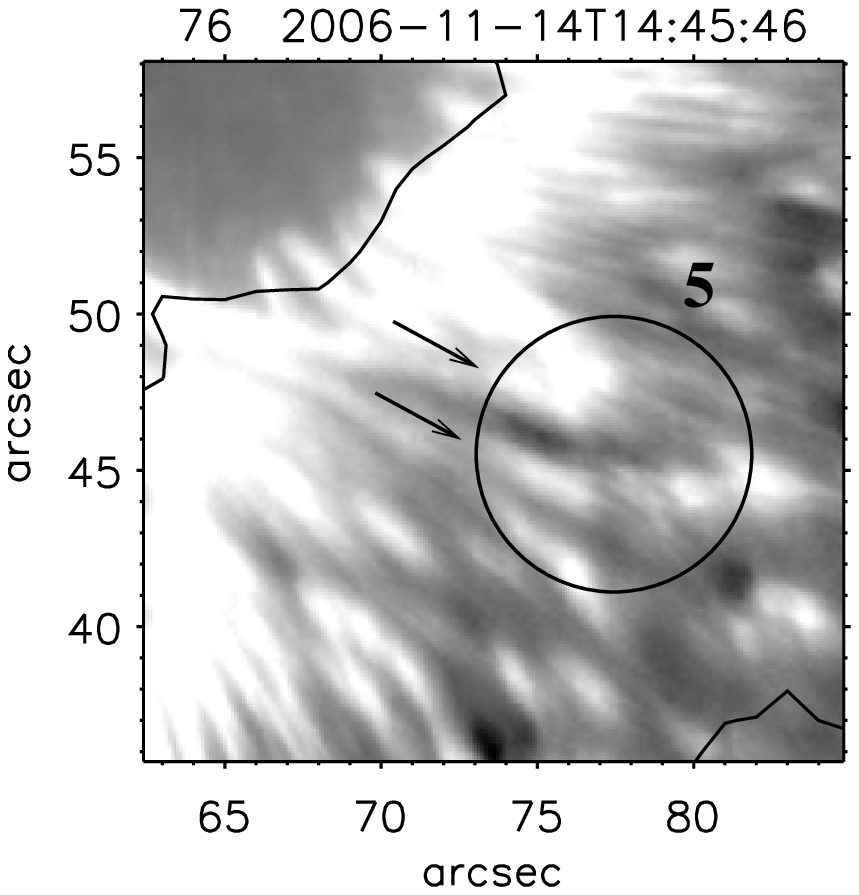}
\includegraphics[height=4.2cm]{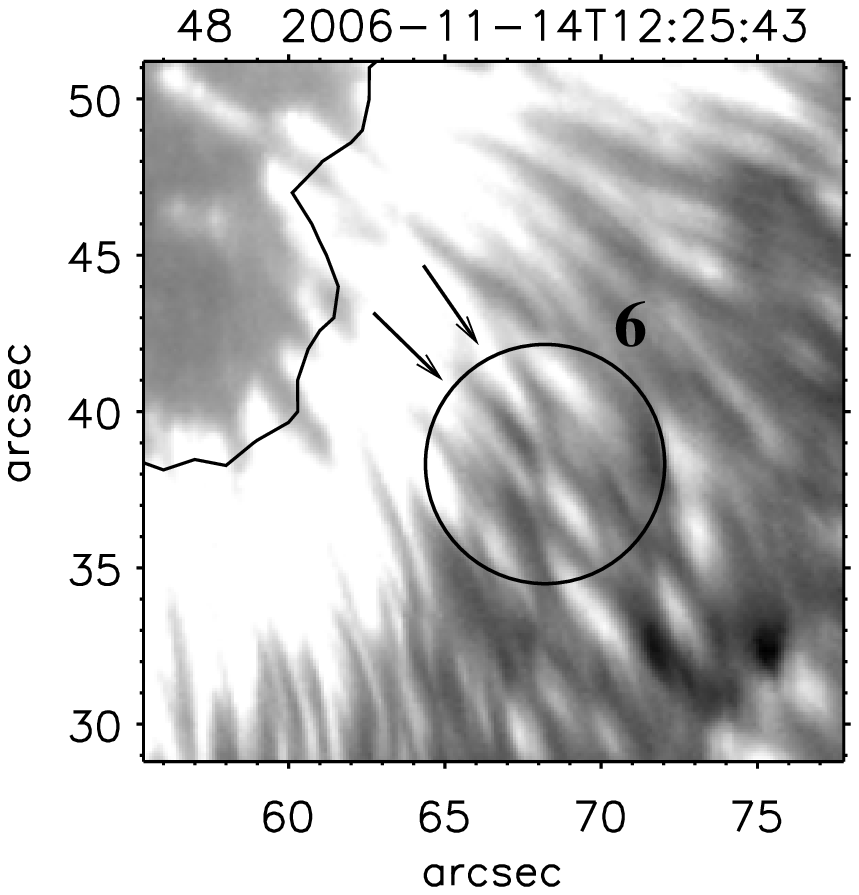}
\includegraphics[height=4.2cm]{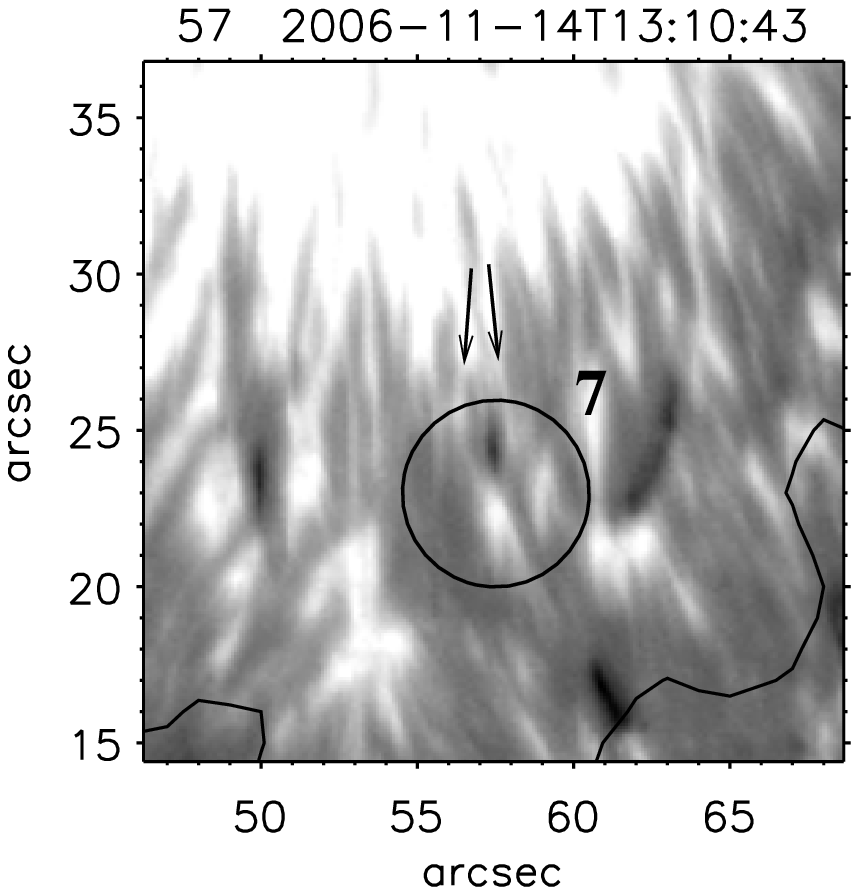}
\includegraphics[height=4.2cm]{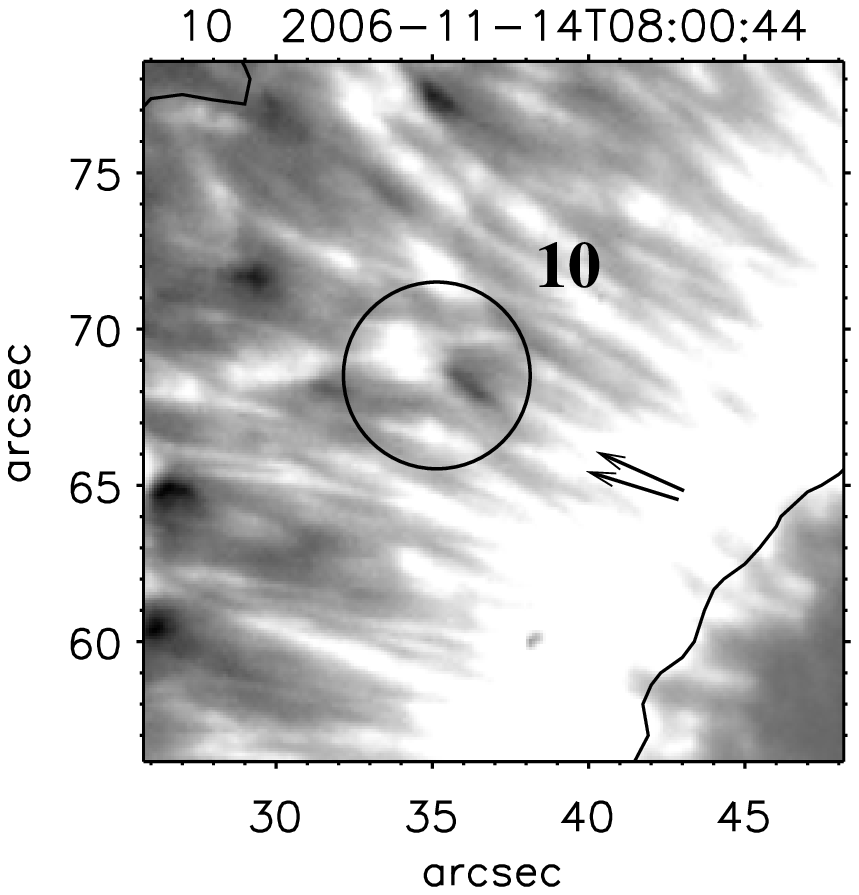}
\includegraphics[height=4.2cm]{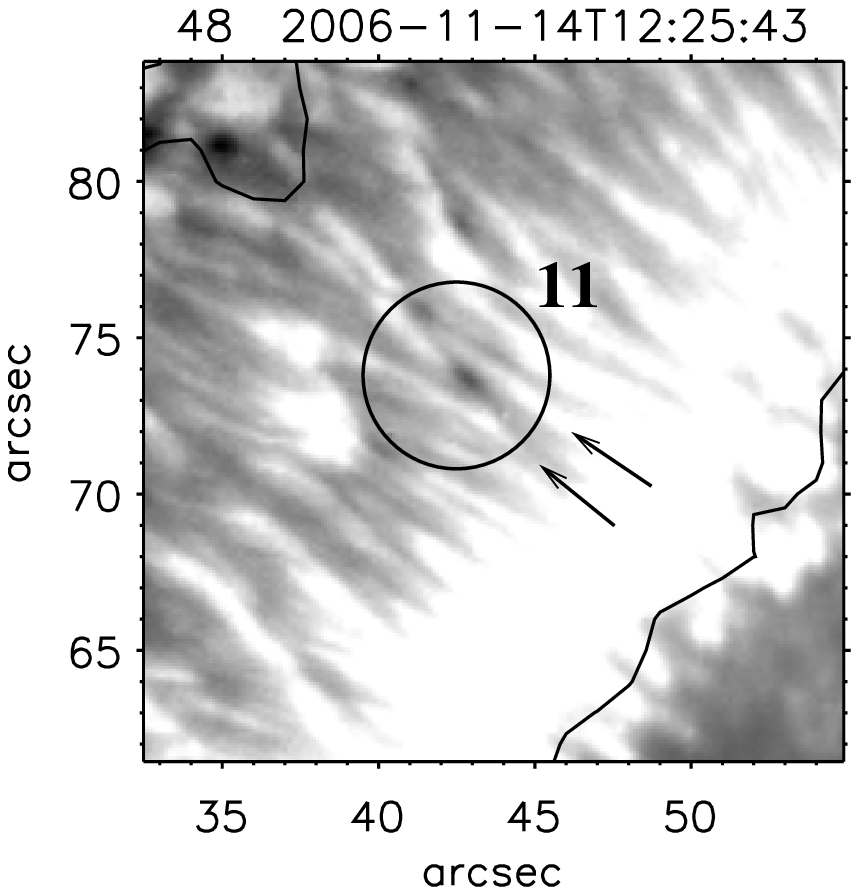}

\caption{{\em Top:} Magnetogram (Stokes $V$ map) of AR 10923 with
umbral and penumbral contours overlaid. The magnetogram signal has
been reversed to treat the spot as if it were of positive
polarity. The direction to disk center is marked with an arrow. The
circles indicate a number of negative-polarity signals in the
penumbra. The magnetogram is saturated at $\pm 500$ 
data numbers (hereafter DNs). {\em Bottom:}
$22\farcs5 \times 22\farcs5$ subfields showing some of the
negative-polarity patches. Usually, they are associated with a
positive-polarity structure farther from the umbra. The bipolar pairs
occur in between filaments of positive polarity (indicated with
arrows). The frame number and time of each magnetogram is given 
in the title. \label{ima_mapas_det1}}

\end{figure*}

\begin{figure*}
\centering
\includegraphics[height=3.8cm, bb=25 0 325 255]{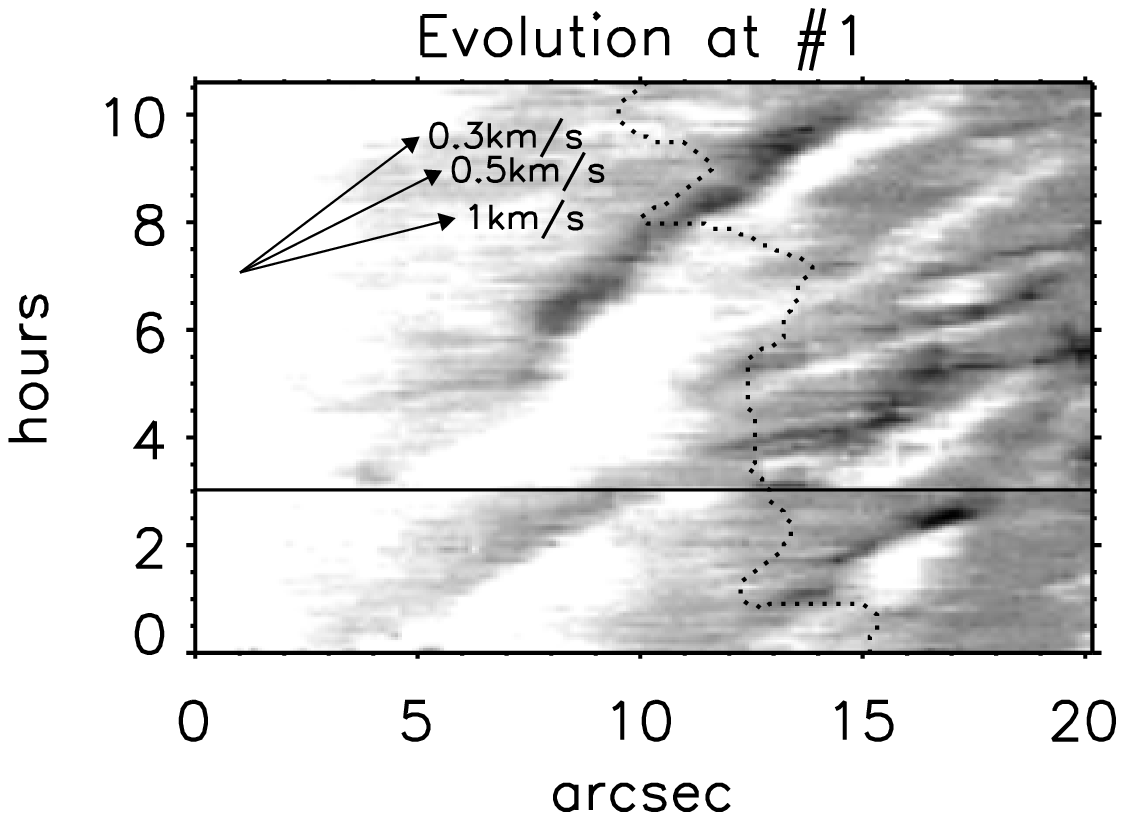}
\includegraphics[height=3.8cm,bb=25 0 325 255,clip]{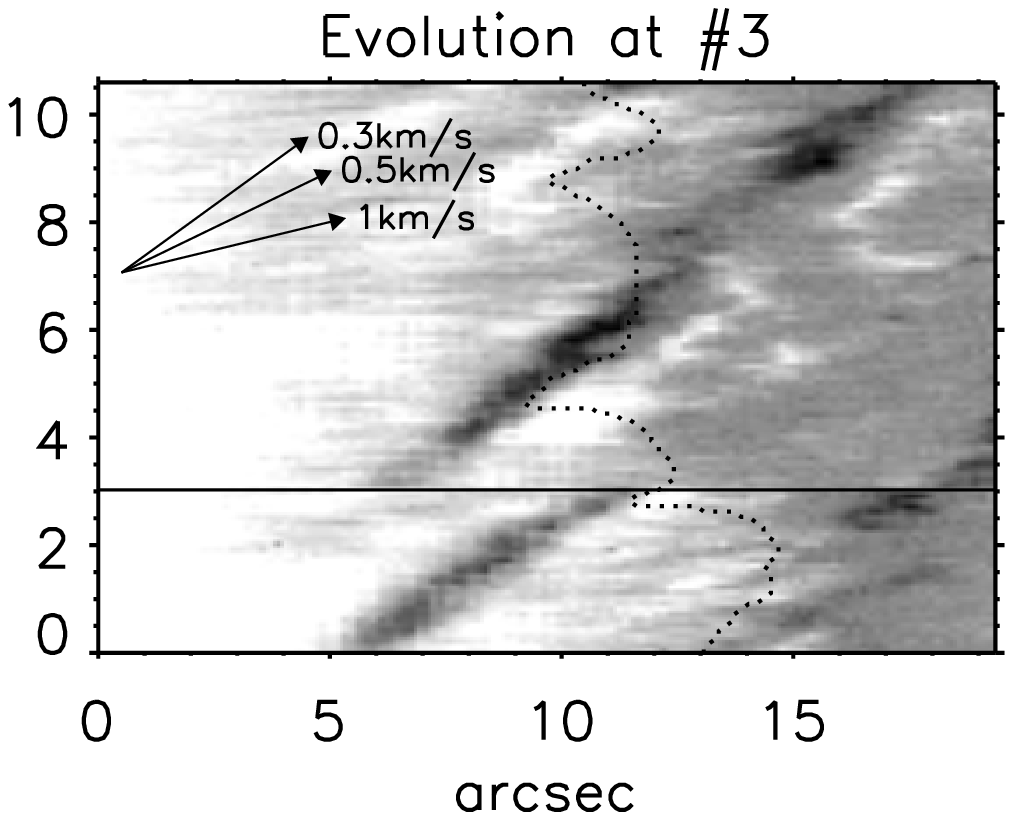}
\includegraphics[height=3.8cm,bb=25 0 325 255,clip]{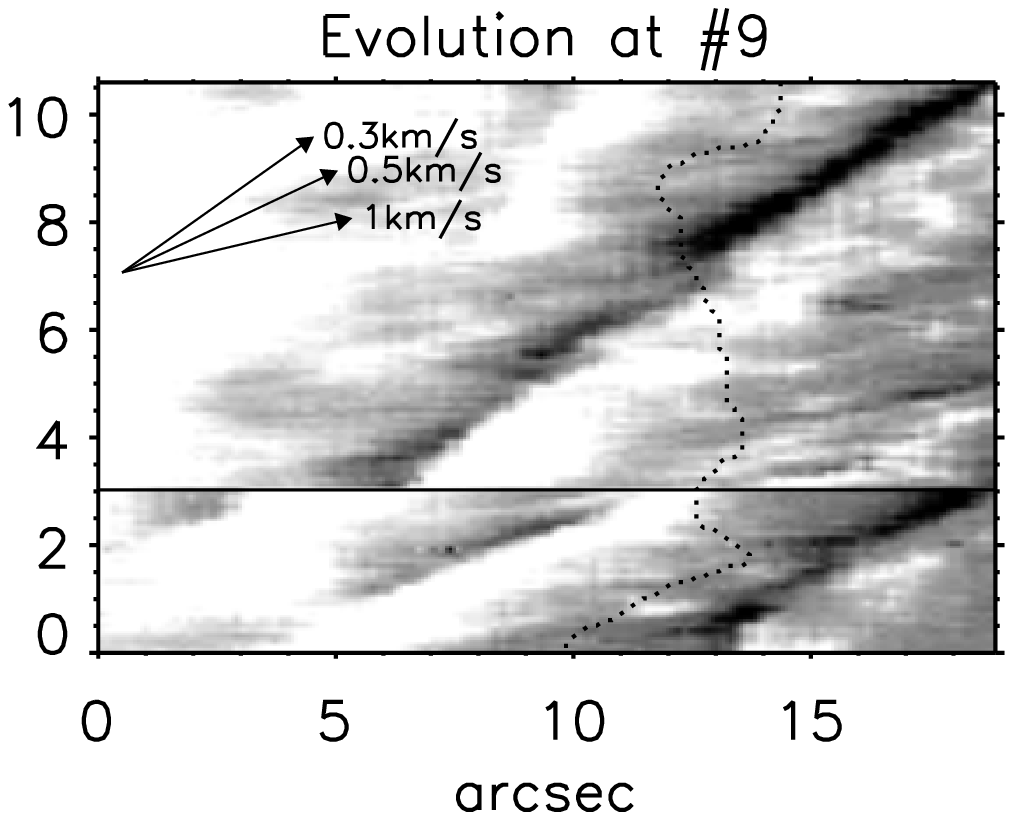}
\includegraphics[height=3.8cm,bb=25 0 325 255,clip]{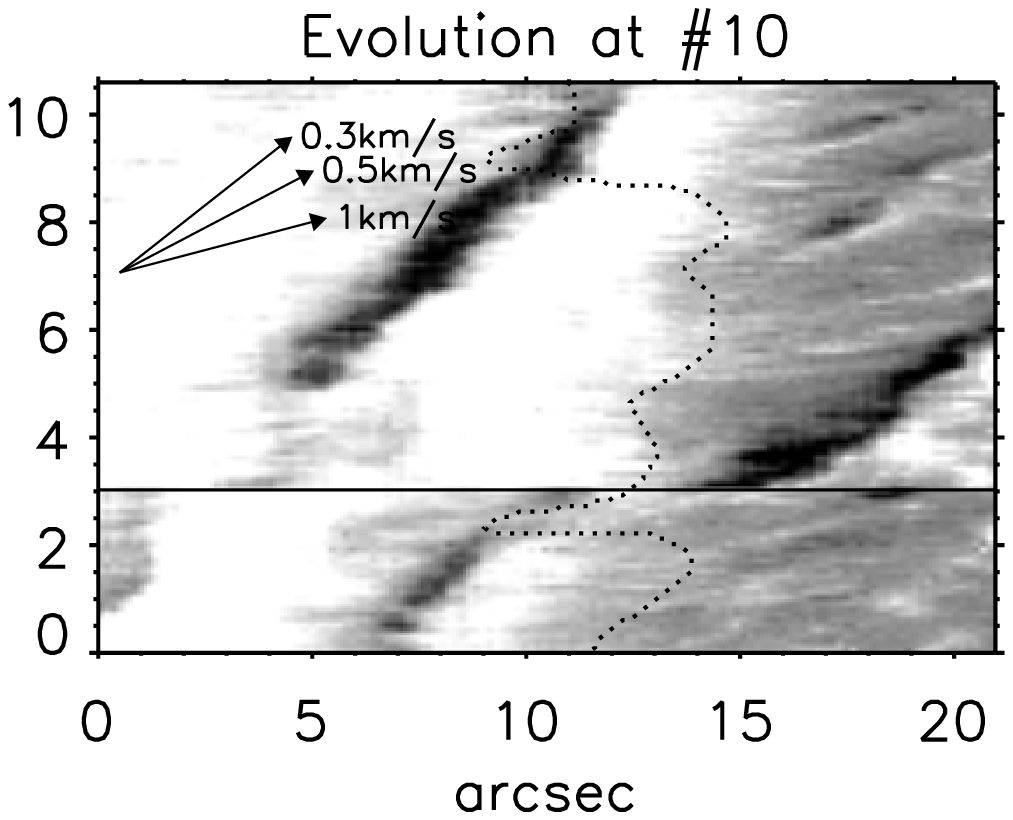}

\caption{Time-slice diagrams for filaments \#1, 3, 9, and 10. They
have been created following the trajectories of the opposite-polarity
patches defining the filaments. The horizontal solid lines mark the
position of the gap between 9:40 and 11:00 UT. The dotted lines
represent the instantaneous outer penumbral boundary.  The
arrows show the slopes of structures moving at 0.3, 0.5, and 1
km~s$^{-1}$. \label{ima_slides_plot}} 
\end{figure*}

Single wavelength magnetograms, such as the ones used here, may fail to
retrieve the polarity of the field in the presence of large Doppler
shifts or multilobed Stokes $V$ spectra.  These profiles are common
in sunspot penumbrae because of their strong Evershed flows 
(e.g., Schlichenmaier \& Collados 2002; Bellot Rubio et al.\ 2007).
However, in sunspots close to disk center the horizontal Evershed flow
does not significantly shift the Stokes $V$ spectra. Only near the
outer edge of the spot, where field lines are thought to return to the
surface, may the flow produce redshifts in both the center-side and 
the limb-side penumbra. Our measurements were taken in the blue wing of
\ion{Fe}{i} 630.25~nm, so Doppler shifts to the red cannot change the
sign of the magnetogram signal.  In addition, the small heliocentric
angle of the spot means that projection effects are negligible.  For
these reasons, negative polarities in the magnetograms truly
correspond to field lines pointing to the solar surface, whereas
positive polarities indicate fields directed away from the sun.

The magnetograms have been aligned by cross-correlation to create a
movie for the 07:10--17:10 UT period with a gap between 09:40 and
11:00 UT ({\tt AR10923.mpg}, available as Supplementary Material). 
This presentation of the data makes it easier to follow the temporal 
evolution of the polarization signals in the penumbra. The movie 
covers a FOV of $112\arcsec\/ \times 112\arcsec$, and has a regular 
cadence of 5 minutes.

\section{Results}

The upper panel of Fig.~\ref{ima_mapas_det1} shows a magnetogram of AR
10923. The polarity of the spot is negative, but for clarity purposes
we have reversed the sign of the magnetograms to treat it as a
positive-polarity spot. With a color scale saturating at $\pm 500$
DNs, the filamentary structure of the penumbra is clearly seen. Also
conspicuous is the existence of many negative (black) polarity patches
well within the penumbra. They are observed not only on the limb side,
but also on the center side of the spot. The positions of some of
these structures are indicated with circles.

A careful inspection of the magnetograms reveals that most of the
negative-polarity patches are associated with positive-polarity
structures located next to them but farther from the umbra. When
detected, the positive-polarity patch has a stronger magnetogram
signal than the background (which is also positive). Usually, the 
two patches appear elongated and narrow. The magnetogram movie shows 
that these opposite-polarity pairs move radially outward as a single
entity. 

The bottom panels of  Fig.~\ref{ima_mapas_det1} give examples of 
bipolar penumbral structures. We have included contours of the
umbra and the penumbra (obtained from the intensity images) to 
help identify the radial distances at which they are seen. 
Figure~\ref{ima_mapas_det1} illustrates the following
properties of these structures: 

\begin{itemize}
\item They appear in the mid penumbra (e.g., \#3, 5, and 10). 
\item  They occur between penumbral filaments with enhanced 
positive signals (see the arrows in the panels of Fig.~\ref{ima_mapas_det1}).  
\item They exhibit a filamentary structure (e.g, \#3, 6, and 7). 
\item The lengths of the individual patches are usually in
the range 2-3\arcsec, although some extreme cases of 5\arcsec\/ 
have been detected. Their mean width is 1\farcs5, but
they can be as narrow as 0\farcs5.  
\end{itemize}

To investigate the temporal evolution of the bipolar structures in
more detail we use time-slice diagrams. They represent the
polarization signals observed along the trajectories of the patches as
a function of time. The width of the paths is 0\farcs16, corresponding
to one pixel in the magnetograms.  Figure~\ref{ima_slides_plot}
displays time-slice diagrams for filaments \#1, 3, 9, and 10. As can
be seen, the typical separation between the two opposite-polarity
patches is on the order of 1-2\arcsec\/. They move with velocities in
the range 0.3-0.4 km~s$^{-1}$, although others propagate as fast as
0.7 km~s$^{-1}$. No significant velocity variations are detected.
Occasionally, more than one bipolar pair seem to travel along the
same filament, one after the other. This appears to be the case in the
four examples of Fig.~\ref{ima_slides_plot}, but we cannot draw 
a definite conclusion due to the gap between 9:40 and 11:00 UT.
 
The time-slice diagrams show that the positive and negative patches
often cross the outer penumbral boundary (represented by the dotted
lines). In the case of filaments \#1 and \#9, the two patches travel a
distance of some 3-6\arcsec\/ in the moat. The crossing of the
penumbral edge does not modify the speed of these features, but they
adopt a more roundish shape outside the spot.

The magnetogram movie demonstrates that, at any given time, up to
15-20 bipolar structures may coexist in the penumbra. They are
observed all around the umbra, with no preferred direction. Their
lifetime can be as short as 30 minutes or longer than 7 hours. 
Usually, the former are very weak and do not reach the outer 
penumbral boundary.

\section{Discussion and conclusions} 

The appearance and coherent motion of the bipolar structures described
above gives a strong impression that they represent the two footpoints
of magnetic field lines having the shape of a sea serpent. The
opposite-polarity patches are observed in regions of weak positive
magnetogram signals, flanked by penumbral filaments with stronger
signals. Since the spot was virtually at disk center, weaker signals
indicate more horizontal and/or weaker fields. These fields are
precisely the ones that carry the Evershed flow. Thus, the bipolar
structures occur in the more inclined fields of the penumbra,
surrounded by the vertical fields of the spines (Lites et al.\ 1993).

We suggest that the opposite-polarity patches are the manifestation of
$\mho$-loop perturbations of the field lines driving the Evershed
flow. The negative-polarity patch, the one closer to the umbra,
contains field lines returning to the solar surface, while the leading
positive-polarity patch represents the re-emergence of the same field
lines. The fact that the magnetogram signal is stronger in the
positive-polarity patch compared with its surroundings indicates more
vertical fields than both the spines and intra-spines, consistent with
the idea that it represents the upstream footpoint of a $\mho$-loop.
There have been reports of negative-polarity field lines from MDI
magnetograms (Sainz Dalda \& Mart\'{\i}nez Pillet 2005; Ravindra
2006), SST magnetograms (Langhans et al.\ 2005), and Hinode
measurements (Ichimoto et al.\ 2007; Bellot Rubio et al.\ 2007), but
this is the first time they are (a) observed in the center-side
penumbra; (b) associated with positive-polarity counterparts that move
together toward the edge of the spot; and (c) identified as the
footpoints of sea-serpent field lines.  Very likely, these structures
are the ones producing the Evershed clouds discovered by Shine et al.\
(1994) and analyzed in detail by Cabrera Solana et al.\ (2007, 2008). The
scenario favored by the present observations is similar to that
resulting from the moving tube simulations of Schlichenmaier (2002). 
In the simulations, penumbral field lines associated with strong 
Evershed flows develop sea-serpent shapes during their rise from 
the spot magnetopause.

\begin{figure}
\centering
\includegraphics[width=8.5cm]{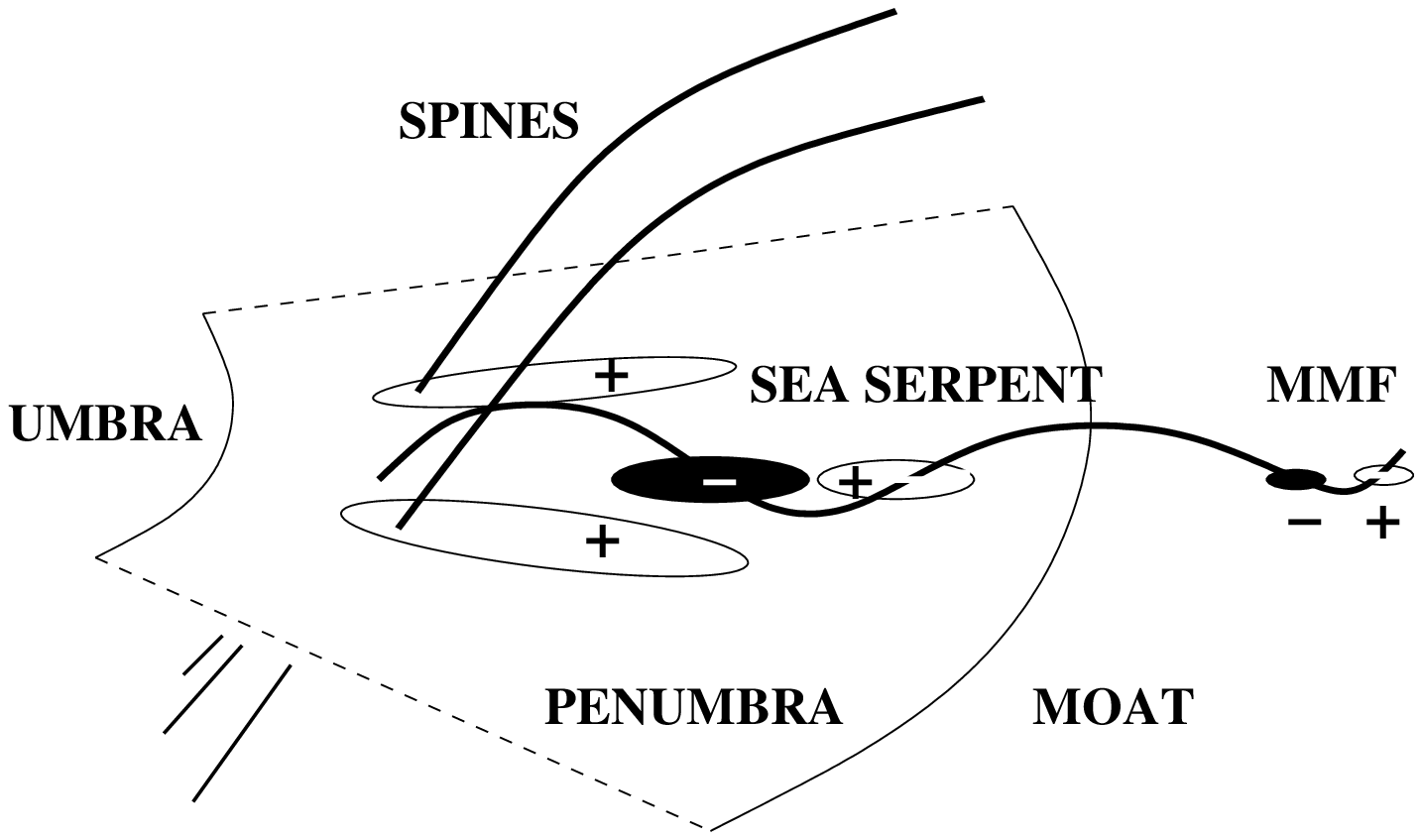}
\caption{Sketch of the penumbra showing sea-serpent field lines
in between more vertical spine fields. The ovals indicate the polarity
of the different structures when observed at disk center.
\label{seaserpents_sketch}} 
\end{figure}

Our measurements show with unprecedented clarity that most of the
opposite-polarity patches reach the edge of the spot and enter the
moat. The transition is smooth, with no changes in the propagation
speed. In other words, these structures are the precursors of the bipolar
moving magnetic features observed around sunspots, confirming the
results of Zhang et al.\ (2003), Sainz Dalda \& Mart\'{\i}nez Pillet
(2005), Cabrera Solana et al.\ (2006), and Kubo et al.\ (2007a,
2007b). MMFs are the continuation of the penumbral fields that harbor
the Evershed flow, as suggested theoretically by Schlichenmaier
(2002), Thomas et al.\ (2002), and others. Their magnetic configuration
must be that of a $\mho$-loop, since this is the shape of the field lines
in the bipolar structures while they still reside in the penumbra. 
Zhang et al.\ (2007) and Cabrera Solana (2007) have deduced a similar
configuration for bipolar MMFs. 

Our interpretation of the observations is summarized in
Fig.~\ref{seaserpents_sketch}.  We draw the inclined fields of the
penumbra as sea serpents flanked by more vertical field lines
representing the penumbral spines.  The sea serpents propagate across
the penumbra and reach the moat, where they become bipolar MMFs.

This magnetic configuration is in good agreement with the moving tube
model of Schlichenmaier et al.\ (1998), which is essentially a siphon
flow (Meyer \& Schmidt 1968; Degenhardt 1989, 1991) with supercritical
velocities. It remains to be seen whether subcritical and/or critical
siphon flows (Thomas \& Montesinos 1993; Montesinos \& Thomas 1997),
as well as other models of the penumbra, such as the turbulent pumping
model (Thomas et al.\ 2002), the gappy penumbral model (Spruit \&
Scharmer 2006; Scharmer \& Spruit 2006), or 3D simulations of sunspot
penumbrae (e.g., Heinemann et al.\ 2007), can also explain the
observations. In any event, it is clear that the detailed topology of
these field lines must be inferred from spectropolarimetric data,
necessarily at the same or better resolution than that provided by
Hinode.

\begin{acknowledgements} We thank all the scientists involved in
the operation of Hinode as Chief Observers for their continuous 
support. Hinode is a Japanese mission developed and launched by
ISAS/JAXA, with NAOJ as domestic partner and NASA and STFC (UK) as
international partners. It is operated by these agencies in
cooperation with ESA and NSC (Norway). This work has been partially
funded by the Spanish MEC through projects ESP2006-13030-C06-01,   
ESP2006-13030-C06-02, AYA2004-05792, and AYA2007-66502.
\end{acknowledgements}

\end{document}